\documentclass[aps,pra,twocolumn,amsmath,showpacs]{revtex4}
\usepackage{bm}
\usepackage{graphicx}
\begin{document}
\setlength{\arraycolsep}{2pt}
\title{Unitary equivalence between ordinary intelligent states and generalized intelligent states}
\author{Hyunchul Nha}
\affiliation{Department of Physics, Texas A \& M University at Qatar, PO Box 23874, Doha, Qatar} 
\date{September 20, 2007}
\begin{abstract}
Ordinary intelligent states (OIS) hold equality in the Heisenberg uncertainty relation involving two noncommuting observables \{$A$, $B$\}, whereas generalized intelligent states (GIS) do so in the more generalized uncertainty relation, the Schr{\"o}dinger-Robertson inequality. In general, OISs form a subset of GISs. However, if there exists a unitary evolution $U$ that transforms the operators \{$A$, $B$\} to a new pair of operators in a {\it rotation} form, it is shown that an arbitrary GIS can be generated by applying the {\it rotation} operator $U$ to a certain OIS. In this sense, the set of OISs is unitarily equivalent to the set of GISs. It is the case, for example, with the su($2$) and the su($1$,$1$) algebra that have been extensively studied particularly in quantum optics. 
When these algebras are represented by two bosonic operators (nondegenerate case), or by a single bosonic operator (degenerate case), the rotation, or pseudo-rotation, operator $U$ corresponds to phase shift, beam splitting, or parametric amplification, depending on two observables \{$A$, $B$\}.

\end{abstract}
\maketitle
\email{phylove00@gmail.com}

\narrowtext
\section{Introduction} 
For a general quantum state, the uncertainties of two noncommuting observables \{$A$, $B$\} cannot be made arbitrarily small at the same time, and the product of them, $\langle(\Delta A)^2\rangle\langle(\Delta B)^2\rangle$, has a certain lower bound. The Heisenberg uncertainty relation (HUR) \cite{Heisenberg}, which is most widely used, provides the bound as 
\begin{eqnarray}
\langle(\Delta A)^2\rangle\langle(\Delta B)^2\rangle\ge\frac{1}{4}|\langle[A,B]\rangle|^2.
\label{eqn:Heisenberg}
\end{eqnarray}
The quantum states that satisfy equality in~(\ref{eqn:Heisenberg}) are usually referred to as minimum-uncertainty states. The lower bound, however, does not always exhibit a true local minimum, e.g., when the commutator $[A,B]$ is not a constant multiple of identity operator. From this perspective, the states holding equality in~(\ref{eqn:Heisenberg}) are generally termed {\it ordinary intelligent states} (OISs) \cite{Aragone,Eberly}. 

On the other hand, the Schr{\"o}dinger-Robertson relation (SRR) generally provides a stronger bound than the HUR \cite{SR1,SR2} as
\begin{eqnarray}
\langle(\Delta A)^2\rangle\langle(\Delta B)^2\rangle\ge\frac{1}{4}|\langle[A,B]\rangle|^2+\langle\Delta A\Delta B\rangle_S^2,
\label{eqn:SR}
\end{eqnarray}
where the covariance $\langle\Delta A\Delta B\rangle_S$ is defined in a symmetric form as
\begin{eqnarray}
\langle\Delta A\Delta B\rangle_S\equiv\frac{1}{2}\langle\Delta A\Delta B+\Delta B\Delta A\rangle.
\end{eqnarray}
The HUR is a special form of the SRR under the condition $\langle\Delta A\Delta B\rangle_S=0$, which is of course not always true. 
The states holding equality in the SRR are termed {\it generalized intelligent states} (GISs) as an analogy to OISs \cite{Trifonov}. 
The OISs and GISs have been extensively studied for many decades and they have attracted a great deal of interest particularly in the context of squeezing \cite{Bergou,Agarwal0,Puri,Gerry,Luis,Brif2,Campos1}. 
More specifically, the intelligent states for the su(2) and su(1,1) algebras were proposed to employ for quantum optical interferometry to achieve the quantum-limited precision in phase measurement 
 \cite{Yurke,Hillery0,Brif1}.

Furthermore, the su(2) and the su(1,1) algebras have recently attracted some renewed interest from the perspective of quantum information theory particularly for the treatment of continuous variables. 
Specifically, the entanglement criteria applicable to non-Gaussian entangled states were derived from those two algebras \cite{Hillery1,Agarwal,nha1}. 
Very recently, it was also shown that the SRR, in conjunction with partial transposition, can generally provide a stronger inequality than the HUR to detect entanglement \cite{nha2}. As an illustration, the entanglement condition derived from the su(2) and su(1,1) algebra was refined to a form invariant with respect to local phase shift in Ref.~\cite{nha2}. 
On an application side, the intelligent states for the su(2) and su(1,1) algebras can be potentially useful for quantum information processing because they all form the class of non-Gaussian entangled states when expressed in terms of two boson operators \cite{nha1}.

Quite obviously, an arbitrary OIS, which has the vanishing covariance $\langle\Delta A\Delta B\rangle_S=0$, is also a GIS, but the converse is not always true. 
Thus, the proposition follows that OISs form a subset of GISs in general \cite{Brif2}. 
In the previous literature, there have been a number of attempts to separately obtain the OISs and the GISs for certain algebras, most prominently, for the su(2) and the su(1,1) algebras. In this paper, we aim at clarifying to some extent the connection between the OISs and the GISs. 
In particular, we consider the case in which there exists a unitary operator $U$ that transforms two operators \{$A$, $B$\} to another pair of operators in a {\it rotation} form. In this case, it is shown that an {\it arbitrary} GIS, $|\Psi\rangle$, can be generated from a certain OIS, $|\Phi\rangle$, by applying the {\it rotation} operator $U$ as $|\Psi\rangle=U|\Phi\rangle$. 
In this sense, it can be said that the set of OISs is unitarily equivalent to the set of GISs. This is particularly the case with the su($2$) and the su($1$,$1$) algebra, of which operators can be represented in terms of boson operators. The unitary operator $U$ is then realized by phase shifter, beam splitter, or parametric amplifier, depending on the pair of two operators \{$A$, $B$\}.

This paper is organized as follows. In Sec.~II, the intelligent states are briefly introduced with their statistical properties. 
In Sec.~III, the equivalence between the set of OISs and that of GISs is demonstrated under the condition that there exists a unitary operation 
that transform the two observables \{$A$, $B$\} to \{$A'$, $B'$\} in a form of rotation. This finding is more concretized for the cases of the su(2) and the su(1,1) algebras in Sec.~IV, 
and the main results are summarized in Sec.~V.

\section{Intelligent states}
First, let us briefly introduce the intelligent states with their statistical characteristics. 
The SRR in Eq.~(\ref{eqn:SR}) can be derived from the Cauchy-Schwartz inequality 
\begin{eqnarray}
\langle f|f\rangle\langle g|g\rangle\ge|\langle f|g\rangle|^2,
\label{eqn:cs}
\end{eqnarray}
where the state vectors $|f\rangle$ and $|g\rangle$ are given by $|f\rangle=\Delta A|\Psi\rangle$ and $|g\rangle=\Delta B|\Psi\rangle$, respectively, for a general state $|\Psi\rangle$\cite{Dodonov1}. 
The variance operator $\Delta O$ is defined as $\Delta O\equiv O-\langle O\rangle$, where $\langle O\rangle$ is the quantum average for the state $|\Psi\rangle$ ($O=A,B$).    

Clearly, the equality holds in Eq.~(\ref{eqn:cs}) when the two vectors are linearly dependent, i.e., $|f\rangle=-i\lambda|g\rangle$, where the parameter $\lambda=\lambda_x+i\lambda_y$ is complex in general. In other words, the GISs, $|\Psi\rangle$, satisfy the characteristic eigenvalue equation
\begin{eqnarray}
(A+i\lambda B)|\Psi\rangle=\beta|\Psi\rangle,
\label{eqn:cegis}
\end{eqnarray}
where $\beta=\langle A\rangle+i\lambda\langle B\rangle$. 

From Eq.~(\ref{eqn:cegis}), the equality in the SRR follows along with the condition $\langle(\Delta A)^2\rangle=|\lambda|^2\langle(\Delta B)^2\rangle$ for a general $\lambda$.  In light of the SRR [Eq.~(\ref{eqn:SR})], coherent states may be defined as those ones for which the two variances, $\langle(\Delta A)^2\rangle$ and $\langle(\Delta B)^2$, are all equal to $V_c\equiv\sqrt{\frac{1}{4}|\langle[A,B]\rangle|^2+\langle\Delta A\Delta B\rangle_S^2}$ ( case of $|\lambda|=1$). On the other hand, squeezing may be defined as one of the two variances reduced below the critical value $V_c$ \cite{Puri}.  In other words, if $|\lambda|$ is smaller (larger) than unity, the observable $A$ ($B$) is squeezed, and the degree of squeezing is parameterized by $|\lambda|$.

{\bf Special cases}: 

(i) if the squeezing parameter $\lambda$ is real ($\lambda_y=0$), the condition $\langle\Delta A\Delta B\rangle_S=0$ follows from Eq.~(\ref{eqn:cegis}), hence the equality in Eq.~(\ref{eqn:Heisenberg}). In other words, the OISs are obtained by solving the eigenvalue equation, Eq.~(\ref{eqn:cegis}), for real values of $\lambda$. 

(ii) On the other hand, if $\lambda$ is pure imaginary ($\lambda_x=0$), it follows that $\langle[A,B]\rangle=0$, and the ordinary Heisenberg uncertainty relation only provides a trivial lower bound, zero \cite{Puri}.

\section{Equivalence between Ordinary Intelligent States and Generalized Intelligent States}
In this section, we consider the connection between the set of OISs and the set of GISs on the condition that the two operators $\{A,B\}$ can be transformed by a certain unitary operator $U$ to new operators $\{A',B'\}$ in a form of rotation. That is,  
\begin{eqnarray}
\begin{pmatrix}&A'\\&B'
\end{pmatrix}
=U\begin{pmatrix}&A\\&B
\end{pmatrix}U^{\dag}
=\begin{pmatrix}
&\cos\phi&-\sin\phi\\&\sin\phi&\cos\phi
\end{pmatrix}
\begin{pmatrix}&A\\&B
\end{pmatrix}.
\label{eqn:rotation}
\end{eqnarray}

(i) First, let the state $|\Phi\rangle$ be an OIS satisfying the eigenvalue equation
\begin{eqnarray}
(A+i\lambda B)|\Phi\rangle=\beta|\Phi\rangle,
\label{eqn:eigeno}
\end{eqnarray}
where $\lambda$ is real \cite{nha0}. 
On applying the unitary operator $U$ on both sides of Eq.~(\ref{eqn:eigeno}), we obtain the eigenvalue equation as 
\begin{eqnarray}
(A+i\Lambda B)|\Psi\rangle=\beta'|\Psi\rangle,
\label{eqn:eigeng}
\end{eqnarray}
where $|\Psi\rangle$ is defined as $|\Psi\rangle\equiv U|\Phi\rangle$. The new parameters $\Lambda$ and $\beta'$ are given by
\begin{eqnarray}
\Lambda&\equiv&\frac{\lambda\cos\phi+i\sin\phi}{\cos\phi+i\lambda\sin\phi},\nonumber\\
\beta'&\equiv&\frac{\beta}{\left(\cos\phi+i\lambda\sin\phi\right)},
\label{eqn:Llrel}
\end{eqnarray}
respectively. 

Note that $\Lambda$ can take any arbitrary complex values in Eq.~(\ref{eqn:Llrel}), and the transformed state $|\Psi\rangle=U|\Phi\rangle$ is thus none other than a certain GIS. 
In other words, for an arbitrarily fixed value of $\Lambda\equiv \Lambda_x+i\Lambda_y$, one can choose the real squeezing parameter $\lambda$ and the rotation angle $\phi$ as
\begin{eqnarray}
\tan2\phi&=&\frac{2\Lambda_y}{1-\Lambda_x^2-\Lambda_y^2},\hspace{1cm}\left(-\frac{\pi}{4}<\phi\le\frac{\pi}{4}\right)\nonumber\\
\lambda&=&\frac{\Lambda_x}{1+\Lambda_y\tan\phi}.
\label{eqn:Llrel1}
\end{eqnarray} 
In short, if their exists a certain unitary operator $U$ that implements the rotation as in Eq.~(\ref{eqn:rotation}), an arbitrary GIS can be generated from a certain OIS by applying the unitary operator $U$, as prescribed in Eq.~(\ref{eqn:Llrel1}). 

(ii) The converse is of course true, and in fact, an OIS is by definition a GIS. Furthermore, an arbitrary GIS can be transformed to an OIS under the same unitary operator $U$. To more deeply understand how it works, let us take a different perspective as follows.  
Consider a $2\times2$ covariance matrix $C$ of which elements are defined as 
\begin{eqnarray}
C_{ij}\equiv\frac{1}{2}\langle\Delta O_i\Delta O_j+\Delta O_j\Delta O_i\rangle,\hspace{0.5cm}(i,j=1,2),
\end{eqnarray}
where $O_1\equiv A$ and $O_2\equiv B$. 
Namely, 
\begin{eqnarray}
C=\begin{pmatrix}
&\langle(\Delta A)^2\rangle&\langle\Delta A\Delta B\rangle_S\\&\langle\Delta A\Delta B\rangle_S&\langle(\Delta B)^2\rangle
\end{pmatrix}.
\label{eqn:matC}
\end{eqnarray}
Then, the determinant of the matrix $C$ is given by 
\begin{eqnarray}
{\rm Det}\{C\}=\langle(\Delta A)^2\rangle\langle(\Delta B)^2\rangle-\langle\Delta A\Delta B\rangle_S^2,
\label{eqn:det}
\end{eqnarray}
which is invariant under rotation in Eq.~(\ref{eqn:rotation}). 
The characteristic equation satisfied by the GISs in Eq.~(\ref{eqn:SR}) now reads as 
\begin{eqnarray}
{\rm Det}\{C\}=\frac{1}{4}|\langle[A,B]\rangle|^2.
\label{eqn:gis}
\end{eqnarray}
Suppose that the state $|\Psi\rangle$ satisfies Eq.~(\ref{eqn:gis}). Then, due to the relation $[A,B]=[A',B']$ and the invariance under rotation, the inversely transformed state $|\Phi\rangle=U^{\dag}|\Psi\rangle$ must also satisfy Eq.~(\ref{eqn:gis}). 
More importantly, the off-diagonal covariance for $|\Phi\rangle$ becomes 
\begin{eqnarray}
\langle\Delta A\Delta B\rangle_{S,|\Phi\rangle}=&&\frac{1}{2}\sin2\phi\left[\langle(\Delta B)^2\rangle-\langle(\Delta A)^2\rangle\right]\nonumber\\&&+\cos2\phi\langle\Delta A\Delta B\rangle_S,
\label{eqn:transcov}
\end{eqnarray} 
by the relation in Eq.~(\ref{eqn:rotation}). 
Note that the quantum averages on the right side of Eq.~(\ref{eqn:transcov}) refer to the ones for the state $|\Psi\rangle$.
Thus, if one chooses the rotation angle $\phi$ as 
\begin{eqnarray}
\tan2\phi=\frac{2\langle\Delta A\Delta B\rangle_S}{\langle(\Delta A)^2\rangle-\langle(\Delta B)^2\rangle},
\label{eqn:pc}
\end{eqnarray} 
the off-diagonal covariance $\langle\Delta A\Delta B\rangle_{S,|\Phi\rangle}$ vanishes in the rotated frame. That is, the GIS, $|\Psi\rangle$, is transformed to an OIS, $|\Phi\rangle$, satisfying 
\begin{eqnarray}
{\rm Det}\{C\}=\langle(\Delta A)^2\rangle\langle(\Delta B)^2\rangle=\frac{1}{4}|\langle[A,B]\rangle|^2,
\end{eqnarray}
under the rotation by the unitary operation $U^{\dag}$.

By (i) and (ii),  the set of OISs is unitarily equivalent to the set of GISs.

\section{su($2$)- and su($1$,$1$)-intelligent states}
In this section, we show that the preceding argument can be generally applied to the su($2$) and the su($1$,$1$) algebras along with their intelligent states. 
\subsection{su(2)-intelligent states}
The su(2) algebra describes the angular momentum operators, as characterized by the commutation relations,
\begin{eqnarray}
\left[J_i,J_j\right]=i\epsilon_{ijk}J_k,\hspace{1cm}(i,j,k=1,2,3),
\end{eqnarray}
where $J_i$'s denote the Cartesian components of the angular momentum. For the choice of $A=J_1$ and $B=J_2$, the relation in~(\ref{eqn:rotation}) can be implemented by the unitary operator $U=e^{i\phi J_3}$. Therefore, a generalized intelligent state $|\Psi\rangle$ can be written in a form as $|\Psi\rangle=e^{i\phi J_3}|\Phi\rangle$, where $|\Phi\rangle$ is an ordinary intelligent state satisfying the eigenvalue equation
\begin{eqnarray}
(J_1+i\lambda J_2)|\Phi\rangle=\beta|\Phi\rangle,
\label{eqn:eigenosu2}
\end{eqnarray}
for a real $\lambda$. 

As a specific example, let us consider the GISs for which the condition $\langle[J_1,J_2]\rangle=i\langle J_3\rangle=0$ holds, i.e., 
the case that the squeezing parameter $\Lambda$ is pure imaginary in Eq.~(\ref{eqn:eigeng}). (See the last paragraph of Sec.~II.) 
Then, with $\Lambda_x=0$ in Eq.~(\ref{eqn:Llrel1}), one has the prescription $\lambda=0$ and $\tan\phi=\Lambda_y$. In other words, we start with the ordinary intelligent state satisfying $J_1|\Phi\rangle=\beta|\Phi\rangle$ in Eq.~(\ref{eqn:eigenosu2}), which is none other than the $J_1$-eigenstate. Since a general $J_1$-eigenstate can be obtained by applying the rotation $e^{-i\frac{\pi}{2}J_2}$ to the $J_3$- eigenstates $|J,m\rangle$, a generalized intelligent state $|\Psi\rangle$ is expressed as $|\Psi\rangle=e^{i\phi J_3}e^{-i\frac{\pi}{2}J_2}|J,m\rangle$. 
This class of intelligent states was in fact studied by R.~Puri, and the expression for those states given in Ref.~\cite{Puri} exactly coincides with that obtained here.

Note that the above argument equally applies to other pairs of observables due to the permutation symmetry in the su(2) algebra. 
For example, for the pair of observables $\{J_2,J_3\}$, the GISs can be obtained by applying the rotation $U=e^{i\phi J_1}$ to the OISs.

In the case that the angular momentum operators are represented by two boson operators $a$ and $b$ \cite{Schwinger}, as 
\begin{eqnarray}
J_1&=&\frac{1}{2}\left(a^\dag b+ab^\dag\right),\nonumber\\ 
J_2&=&\frac{1}{2i}\left(a^\dag b-ab^\dag\right),\nonumber\\ 
J_3&=&\frac{1}{2}\left(a^\dag a-b^\dag b\right),
\label{eqn:su2operators}
\end{eqnarray}
the unitary operator $e^{i\phi J_3}$ simply denotes a local phase shift for the two modes $a$ and $b$. In fact, only one local phase shift can implement the necessary rotation by a proper choice of phase angle.  On the other hand,  the unitary operators $e^{i\phi J_1}$ and  $e^{i\phi J_2}$ correspond to the action of the beam splitter \cite{Campos2}.  

\subsection{su(1,1)-intelligent states}
In the su(1,1) algebra, the operators $K_x,K_y$ and $K_z$ satisfy the commutation relations, $\left[K_1,K_2\right]=-iK_3,\left[K_2,K_3\right]=iK_1$, and $\left[K_3,K_1\right]=iK_2$. 
In spite that the commutators in the su(1,1) algebra differ in sign from those in the su(2) algebra, the rotation of the operators $K_1$ and $K_2$ can be realized by the unitary operator $e^{i\phi K_3}$, similar to the case in su(2) algebra. That is, the relation $|\Psi\rangle=e^{i\phi K_3}|\Phi\rangle$ holds between GISs and OISs.

Of course, a significant difference can arise in the su(1,1) algebra due to the lack of permutation symmetry. For instance, for a different choice of two observables $\{K_1,K_3\}$, the unitary operator $e^{i\phi K_2}$ does not effect rotation, but gives the transformation as
\begin{eqnarray}
\begin{pmatrix}&K_1'\\&K_3'
\end{pmatrix}
=\begin{pmatrix}
&\cosh\phi&-\sinh\phi\\&-\sinh\phi&\cosh\phi
\end{pmatrix}
\begin{pmatrix}&K_1\\&K_3
\end{pmatrix}.
\label{eqn:hyperrotation}
\end{eqnarray}
Nonetheless, the equivalence of GIS and OIS is similarly deduced along with the lines in Sec.~III. More concretely, the relations in Eq.~(\ref{eqn:Llrel1}) now become
\begin{eqnarray}
\tanh2\phi&=&\frac{2\Lambda_y}{1+\Lambda_x^2+\Lambda_y^2},\hspace{1cm}\nonumber\\
\lambda&=&\frac{\Lambda_x}{1-\Lambda_y\tanh\phi},
\label{eqn:Llrel2}
\end{eqnarray} 
so that the prescriptions for $\phi$ and $\lambda$ exists for any values of $\Lambda_x$ and $\Lambda_y$ to produce an arbitrary GIS from an OIS. 

Conversely, in a similar method used to derive Eqs.~(\ref{eqn:transcov}) and~(\ref{eqn:pc}) in Sec.~III, 
the off-diagonal covariance $\langle\Delta A\Delta B\rangle_{S,|\Phi\rangle}$ can be made vanish by choosing the transformation as
\begin{eqnarray}
\tanh2\phi=\frac{2\langle\Delta A\Delta B\rangle_S}{\langle(\Delta A)^2\rangle+\langle(\Delta B)^2\rangle},
\label{eqn:pcc}
\end{eqnarray} 
to transform a GIS to an OIS.

The su(1,1) operators can be represented by two bosonic operators as
\begin{eqnarray}
K_1&=&\frac{1}{2}\left(a^\dag b^\dag+ab\right),\nonumber\\ 
K_2&=&\frac{1}{2i}\left(a^\dag b^\dag-ab\right),\nonumber\\ 
K_3&=&\frac{1}{2}\left(a^\dag a+b^\dag b+1\right),
\label{eqn:su11operators}
\end{eqnarray}
or by a single bosonic operator as
\begin{eqnarray}
K_1&=&\frac{1}{4}\left(a^{\dag2}+a^2\right),\nonumber\\ 
K_2&=&\frac{1}{4i}\left(a^{\dag2}-a^2\right),\nonumber\\ 
K_3&=&\frac{1}{4}\left(2a^\dag a+1\right).
\label{eqn:su11operators1}
\end{eqnarray}
The unitary operation $e^{i\phi K_3}$ can also be implemented by a local phase shift in both representations. 
On the other hand, $e^{i\phi K_1}$ and $e^{i\phi K_2}$ are realized by the nondegenerate parametric amplification for two-mode case or by the degenerate parametric amplification for single-mode case. 
\section{Summary}
In this paper, the connection of the OISs and the GISs holding equality in the uncertainty relations has been studied and made clarified to some degree. 
In particular, it has been shown that there exists a unitary equivalence between the set of OISs and that of GISs for two noncommuting observables \{$A$, $B$\} in the case that there exists a {\it rotational} unitary operator $U$ for those observables in view of Eq.~(\ref{eqn:rotation}). 
This is particularly true for the su(2) and the su(1,1) algebras, and in the latter case, although only a pseudo-rotation is effected for a particular choice of two observables, 
it was shown that the unitary equivalence still holds good. In the case that these algebras are represented by bosonic operators, the unitary operation corresponds to phase shift, beam splitting, or parametric amplification depending on the choice of the observables.
 


\begin{references}
\bibitem{Heisenberg} W.~Heisenberg, Z.~Phys. {\bf 43}, 122 (1927).
\bibitem{Aragone} C.~Aragone, G.~Guerri, S.~Salamo, and J.~L.~Tani, J.~Phys.~A {\bf 7}, L149 (1974); 
C.~Aragone, E.~Chalbaud, and S.~Salamo, J.~Math.~Phys. {\bf 17}, 1963 (1976).
\bibitem{Eberly} K.~Wodkiewicz and J.~H.~Eberly, J.~Opt.~Soc.~Am.~B {\bf 2}, 458 (1985).
\bibitem{SR1} E.~Schr{\"o}dinger, Sitzunsber. Preuss.~Akad.~Wiss. p.~296 (Berlin,1930).
\bibitem{SR2} H.~R.~Robertson, Phys. Rev. {\bf 46} 794 (1934).
\bibitem{Trifonov} D.~A.~Trifonov, J.~Math.~Phys. {\bf 35}, 2297 (1994).
\bibitem{Bergou} J.~A.~Bergou, M.~Hillery and D.~Yu, \pra {\bf 43}, 515 (1991); D.~Yu and M.~Hillery, Quantum Opt. {\bf 6}, 37 (1994).
\bibitem{Agarwal0} G.~S.~Agarwal and R.~R.~Puri, \pra {\bf 41}, 3782 (1990).
\bibitem{Puri} R.~R.~Puri, \pra {\bf 49}, 2178 (1994).
\bibitem{Gerry} C.~C.~Gerry and R.~Grobe, \pra {\bf 51}, 4123 (1995).
\bibitem{Luis} A.~Luis and J.~Perina, \pra {\bf 53}, 1886 (1996).
\bibitem{Brif2} C.~Brif, Int.~J.~Theor.~Phys. {\bf 36}, 1651 (1997).
\bibitem{Campos1} R.~A.~Campos and C.~G.~Gerry, \pra {\bf 60}, 1572 (1999).
\bibitem{Yurke} B.~Yurke, S.~L.~McCall, and J.~R.~Klauder, \pra {\bf33}, 4033 (1986).
\bibitem{Hillery0} M.~Hillery and L.~Mlodinow, \pra {\bf 48}, 1548 (1993).
\bibitem{Brif1} C.~Brif and A.~Mann, \pra {\bf 54}, 4505 (1996).
\bibitem{Hillery1} M.~Hillery and M.~Zubairy, \prl {\bf 96}, 050503 (2006); M.~Hillery and M.~Zubairy, \pra {\bf 74}, 032333 (2006).
\bibitem{Agarwal} G.~S.~Agarwal and A.~Biswas, New J.~Phys. {\bf 7}, 211 (2005).
\bibitem{nha1} H.~Nha and J.~Kim, \pra {\bf 74}, 012317 (2006).
\bibitem{nha2} H.~Nha, \pra {\bf 76}, 014305 (2007).
\bibitem{Dodonov1} V.~V.~Dodonov, E.~V.~Kurmyshev, and V.~I.~Man'ko, Phys. Lett. {\bf 79}A, 150 (1980); B.~Nagel, eprint quant-ph/9711028. 
\bibitem{nha0} Throughout this paper, the notation $|\Psi\rangle$ represents a GIS, whereas $|\Phi\rangle$ an OIS.
\bibitem{Schwinger} J.~Schwinger, in {\it Quantum Theory of Angular Momentum}, edited by L.~C.~Biedenharn and H.~van~Dam (Academic, New York, 1965).
\bibitem{Campos2} R.~A.~Campos, B.~E.~A.~Saleh, and M.~C.~Teich, \pra {\bf 40}, 1371 (1989).

\end{references}
\end{document}